\documentstyle[psfig,twocolumn,aps]{revtex}
\begin{document}
\draft

%
\twocolumn[\hsize\textwidth\columnwidth\hsize\csname@twocolumnfalse\endcsname
%
%

\title{Specific Heat of the 2D Hubbard Model}

\author{ Daniel Duffy and Adriana Moreo}

\address{Department of Physics, National High Magnetic Field Lab and
MARTECH, Florida State University, Tallahassee, FL 32306, USA}

\date{\today}
\maketitle

\begin{abstract}
Quantum Monte Carlo results for the specific heat $c$ of the two
dimensional Hubbard model are presented.  At half-filling it was
observed that $c \sim T^2$ at very low temperatures. Two distinct
features were also identified: a low temperature peak related to the
spin degrees of freedom and a higher temperature broad peak related to
the charge degrees of freedom. Away from half-filling the spin induced
feature slowly disappears as a function of hole doping while the charge
feature moves to lower temperature.  A comparison with experimental
results for the high temperature cuprates is discussed.
\end{abstract}

\vskip2pc]
\narrowtext

\section{Introduction}

The Hubbard model is among the simplest Hamiltonians that describe the
behavior of correlated electrons. Specially since the discovery of high
temperature superconducting materials, considerable attention has been
devoted to this model and significant progress was achieved in
understanding its ground state properties, particularly at half-filling,
although superconductivity is still elusive\cite{rev}.  Static and
dynamical spin correlations, the optical conductivity and other
observables have been studied in detail.\cite{rev} However, not much
attention has been devoted to its thermodynamical properties despite the
large amount of experimental specific heat measurements available for
the cuprates.  The aim of this paper is to fill that void and to present
a systematic study of the specific heat of the two dimensional (2D)
Hubbard model for different couplings $U/t$, dopings and temperatures.
To achieve that goal Quantum Monte Carlo (QMC) techniques are used.

The Hubbard Hamiltonian is given by
$$
{ H=}
-t{ \sum_{\langle {\bf{ij}} \rangle,\sigma}(c^{\dagger}_{{\bf{i}},\sigma}
c_{{\bf{j}},\sigma}+h.c.)}
$$

$$
+U{ \sum_{{\bf{i}}}(n_{{\bf{i}} \uparrow}-1/2)( n_{{\bf{i}}
\downarrow}-1/2)+\mu\sum_{{\bf{i}},\sigma}n_{{\bf{i}}\sigma} },
\eqno(1)
$$

\noindent where ${ c^{\dagger}_{{\bf{i}},\sigma} }$ creates an electron
at site ${ {\bf i } }$ with spin projection $\sigma$, ${
n_{{\bf{i}}\sigma} }$ is the number operator, the sum ${ \langle
{\bf{ij}} \rangle }$ runs over pairs of nearest neighbor lattice sites,
$U$ is the on-site Coulombic repulsion, ${t}$ the nearest neighbor
hopping amplitude, and $\mu$ the chemical potential. In the following
$t=1$ will be used as the unit of energy. The boundary conditions are
periodic.

\section{Half-filling}

The computational calculation of the specific heat $c$ is not simple.
In principle, $c$ is given by the derivative of the energy $E$ (defined
as $E = \langle H \rangle/N$, with $N$ being the number of sites) with
respect to the temperature $T$ at constant density.  However, note that
in determinantal QMC simulations, which are set up in the grand
canonical ensemble, the energy is a function of the chemical potential
that has to be adjusted to keep the density $\langle n \rangle$ constant
as the temperature changes.  In other words, $\partial E/\partial T$
must be calculated along lines of constant $\langle n \rangle$ in the
$T-\mu$ plane.  In this framework the calculation of $c$ cannot proceed
using $c \sim \langle H^2 \rangle - \langle H \rangle^2$, as when the
number of particles is fixed.  Another detail that is important is the
finite discretization of the derivatives along the lines of constant
density.  Naively, the ratio $\Delta E/\Delta T$, with $\Delta T$ very
small, should be calculated.  However, us-

\begin{figure}[h]
\centerline{\psfig{figure=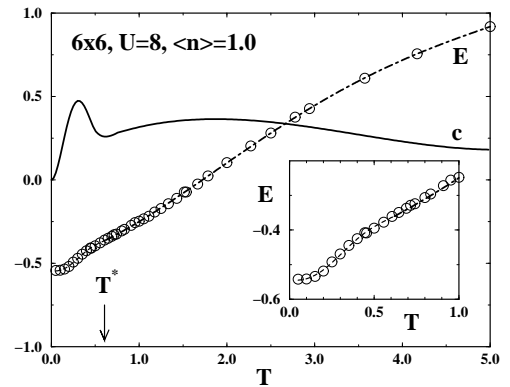,width=7cm,bbllx=98pt,bblly=36pt,bburx=590pt,bbury=700pt,angle=270}}
\caption{Monte Carlo results for the energy $E$ on a $6 \times 6$
cluster at half-filling and $U=8$ (open circles).  The low temperature
polynomial fit is indicated by the dashed line, while the short-long
dashed line indicates the high temperature fit. The solid line denotes
the specific heat $c$. The low temperature data that produce the spin
peak are shown in the inset.  The error bars are smaller than the size
of the dots.}
\label{fig1}
\end{figure}

\begin{figure}[h]
\centerline{\psfig{figure=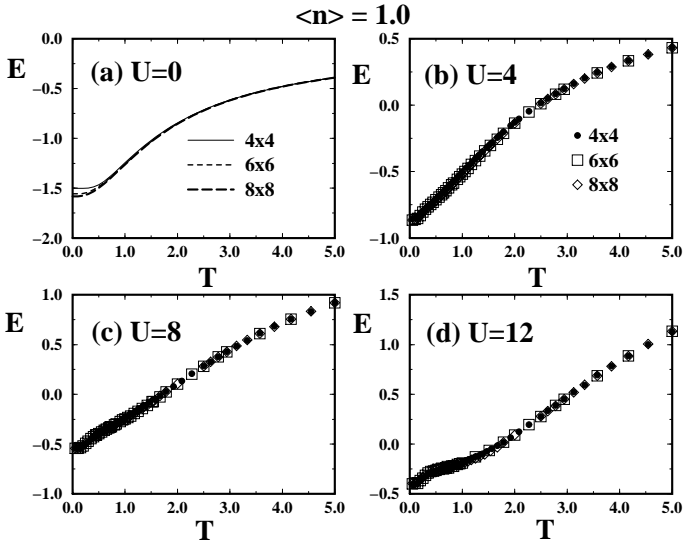,width=9cm,bbllx=60pt,bblly=36pt,bburx=590pt,bbury=750pt,angle=270}}
\caption{Energy $E$ as a function of temperature $T$ at half-filling on
$4 \times 4$, $6 \times 6$ and $8 \times 8$ clusters for a) $U=0$, b)
$U=4$, c) $U=8$ and d) $U=12$. The error bars are smaller than the size
of the dots.}
\label{fig2}
\end{figure}

\noindent ing such a procedure the small statistical error in $E$
introduces large errors in $c$. For that reason we have decided to
calculate $E(T)$ (at fixed $\langle n \rangle$) numerically as
accurately as possible, and then fit the Monte Carlo points with a
polynomial that smears out the small fluctuations in $E$.  $c$ is
obtained by taking derivatives from this polynomial analytically.
Motivated by the shape of the $E$ vs $T$ curve, different polynomials
were used for the high and low temperature regimes. In Fig.1, the raw
Monte Carlo data for $E$ as a function of temperature corresponding to
$U=8$ at half-filling on a $6 \times 6$ cluster are presented. Each data
point was obtained by performing around 10,000 measurement sweeps.  The
dashed line indicates the low temperature fit by a polynomial of order 6
in $T$, while the short-long dashed line indicates the high temperature
fit, in this case to a polynomial of order 4. $T^*$ is the temp-

\begin{figure}[h]
\centerline{\psfig{figure=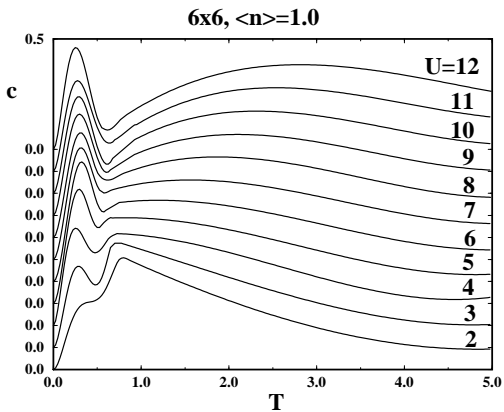,width=7cm,bbllx=60pt,bblly=36pt,bburx=590pt,bbury=700pt,angle=270}}
\caption{Specific heat $c$ vs $T$ at half-filling for different values
of $U$ ranging from 2 to 12. The vertical axis for each coupling is
shifted for clarity.}
\label{fig3}
\end{figure}

\begin{figure}[h]
\centerline{\psfig{figure=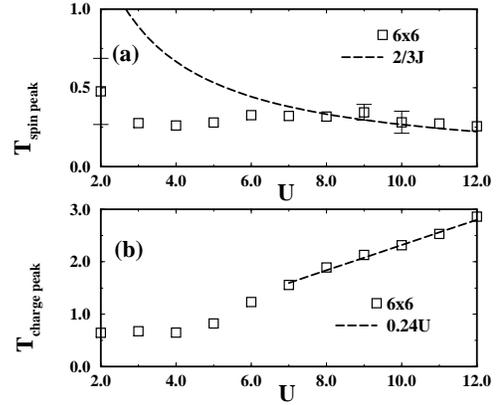,width=7cm,bbllx=98pt,bblly=36pt,bburx=590pt,bbury=700pt,angle=270}}
\caption{a) Temperature $T_{spin~peak}$, where the spin peak is located,
as a function of $U$ at half-filling. The dashed line indicates $T=2J/3$
(asymptotic result in the Heisenberg limit); b) Temperature
$T_{charge~peak}$, where the charge peak is located, as a function of
$U$. The dashed line indicates $T=0.24U$.}
\label{fig4}
\end{figure}

\noindent erature where the two fits meet. Its value depends on the
parameters $U$ and $\langle n \rangle $ and is typically of the order of
1. In order to make a smooth connection of the two fits we included
points below (above) $T^*$ for the high (low) temperature fit within a
window $\sim 0.2$ centered at $T^*$.  The specific heat was obtained
through the analytic derivative of the fitting polynomials, and it is
also shown in Fig.1 with a continuous line.  The inset of the figure
shows with more detail the low energy data that generate the low
temperature peak in $c$ (to be discussed later).

An important issue in QMC simulations are finite size effects (FSE).
Upon studying $4\times 4$, $6\times 6$ and $8\times 8$ clusters, it was
observed that the FSE in $E$ vs $T$ are strong at very weak coupling but
become negligible for $U=8$ or larger.  In Fig.2, the energy of the
different clusters for $U=0$, 4, 8 and 12 is shown.  Since the FSE are
small we decided that results on $6 \times 6$ clusters are
representative of the physical behavior analyzed in this study and,
thus, this is the lattice size that we have used in the remaining of the
paper.  In Fig.3, $c$ vs $T$ at half-filling for different values of $U$
is shown. There are two important features in these curves: 1) A low
temperature peak that appears when the low lying spin states are
excited, and 2) a higher temperature peak which appears when states in
the upper Hubbard band are excited.  In the weak coupling regime the low
temperature peak moves to slightly higher temperature as $U$ increases,
reaching a turning point at $U\approx 7$ where the peak is at $T=0.3$.
For $U > 7$ the peak slowly moves to lower temperatures, as $U$ grows.
This indicates the beginning of the strong coupling regime since it is
well known that for large values of $U$ the Hubbard and the $t-J$ models
have similar behaviors and the coupling constants are related through
$J=4t^2/U$. Numerical studies on the $t-J$ model have indicated that at
half-filling (Heisenberg limit) the peak in $c$ appears at $T\approx
2J/3$\cite{heis} which in terms of $U$ corresponds to $T\approx
8t^2/3U$. Thus, when this regime is reached we expect the peak to move
to lower temperature with increasing $U$.  The position of the peak as a
function of

\begin{figure}[h]
\centerline{\psfig{figure=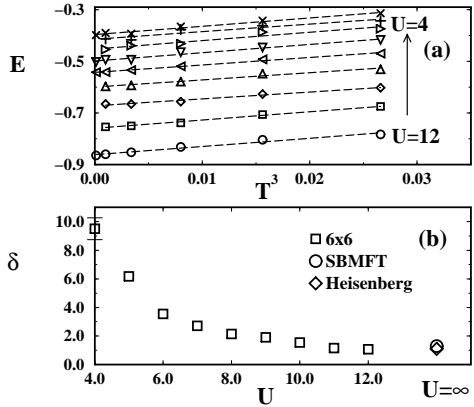,width=7cm,bbllx=98pt,bblly=36pt,bburx=590pt,bbury=700pt,angle=270}}
\caption{a) Energy vs $T^3$ at half-filling for different values of $U$;
b) Coefficient $\delta$ of the low temperature fit $c \sim \delta T^2$,
as a function of $U$ at half-filling. Numerical and analytical values
for the Heisenberg limit ($U=\infty$) are indicated. The circle
corresponds to the mean field result of Ref.[6], and the diamond
denotes the numerical result of Ref.[8].}
\label{fig5}
\end{figure}

\noindent $U/t$ is shown in Fig.4.a, where the dashed line indicates
$T=2J/3$. The asymptotic behavior is reached for $U\geq 10$.

The broad high temperature peak moves to higher temperature as $U$
increases as expected since its presence corresponds to the excitation
of states across the gap that grows with $U$. In Fig.4.b the position of
this peak is shown as a function of $U$. For $U\geq 7$ the dependence of
the position of the peak with $U$ becomes approximately linear, and it
is given by $0.24 U$. A spin-density-wave mean field calculation
 of the gap $\Delta$ as a function of $U$, at large $U$, gives the
result $\Delta \sim 0.48 U$. Apparently, quantum fluctuations reduce the
size of the gap.  Note that in Fig. 3 it can be observed that the
minimum in $c$ between the two peaks becomes deeper as $U$ increases and
the charge peak increases its width.

In previous work the specific heat for the half-filled Hubbard model in
one dimension has been evaluated.\cite{schul,usuki,klumper} We found
that the qualitative behavior in two and one dimensions is similar
regarding the existence and coupling dependence of the two peaks.
However, the following differences were observed: 1) According to
Ref.\cite{schul} the two peaks can be resolved for $U>4$ while here
 we were able to identify the two peaks already at $U=2$.  The fact that
only one maximum is observed in Ref.\cite{schul} in the strong coupling
regime is due to the small $T$ interval considered in their study; 2)
According to Ref.\cite{usuki} the maximum in $c$ associated with the
spin excitations moves to lower temperatures as $U$ increases in weak
coupling while in our 2D study the opposite behavior was found.

Another important feature observed here at half-filling is that at low
temperatures the specific heat follows $c=\delta T^2$, i.e., the
behavior predicted by spin-wave calculations.\cite{arov} In Fig.5.a we
show the energy as a function of $T^3$ for different values of $U$
showing that linear behavior occurs for $T\leq 0.3$.  The value of
$\delta$ depends on $U$, and it decreases as the coupling increases. For
large $U$ the

\begin{figure}[h]
\centerline{\psfig{figure=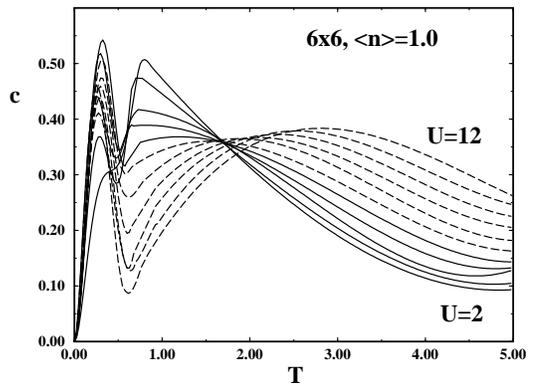,width=7cm,bbllx=98pt,bblly=36pt,bburx=590pt,bbury=700pt,angle=270}}
\caption{$c$ vs $T$ for $U=2$ to 5 (continuous lines) and for $U=6$ to
12 (dashed lines).  All the curves intersect at $T_2\approx 1.6$, while
those corresponding to weak coupling (continuous lines) also intersect
at $T_1\approx 0.6$.}
\label{fig6}
\end{figure}

\noindent limiting value $\delta\approx 1.1$ is obtained in good
agreement with the reported value for the Heisenberg
model.\cite{arov,ted,japan} A slave boson mean field theory (SBMFT)
calculation provided a value of $\delta=1.3\pm 0.05$\cite{arov} while a
numerical study obtained $\delta=1.1\pm 0.2$.\cite{japan} The behavior of
$\delta$ vs $U$ is shown in Fig.5.b.

In Fig.6, $c$ vs $T$ for several values of $U$ ranging from 2 to 12 are
presented. These are the same curves that were shown in Fig.3 but now
using common vertical units.  It is interesting to observe that all the
curves intersect at $T=1.6\pm 0.2$. If only small values of the coupling
are considered, i.e. $U$ ranging from 2 to 5, the curves cross also at
$T_1=0.6$ in addition to $T_2=1.6$.  This behavior was predicted by
Vollhardt\cite{vol} and was observed in the paramagnetic phase of the
infinite dimensional Hubbard model for $0\leq U\leq 2.5$.

\section{Finite Hole Density}

To compare our results with those of the superconducting cuprates it is
important to study the specific heat as 

\begin{figure}[h]
\centerline{\psfig{figure=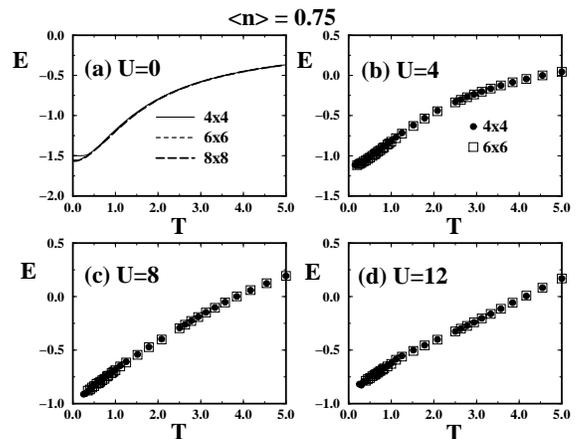,width=7cm,bbllx=60pt,bblly=36pt,bburx=590pt,bbury=700pt,angle=270}}
\caption{Energy $E$ as a function of temperature $T$ on $4 \times 4$ and
$6 \times 6$ clusters and density $\langle n \rangle = 0.75$ for a)
$U=0$, b) $U=4$, c) $U=8$ and d) $U=12$.}
\label{fig7}
\end{figure}

\begin{figure}[h]
\centerline{\psfig{figure=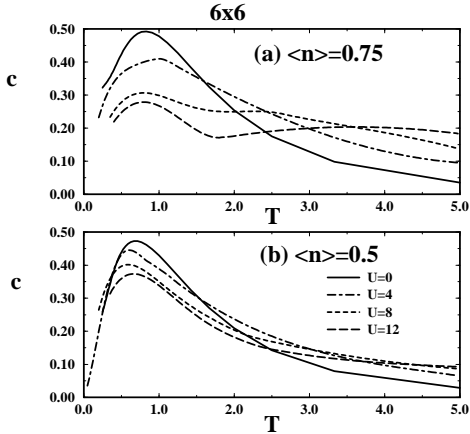,width=7cm,bbllx=60pt,bblly=36pt,bburx=590pt,bbury=700pt,angle=270}}
\caption{$c$ vs $T$ at density a) $\langle n \rangle =0.75$ and b)
$\langle n \rangle =0.50$ for $U=0$, 4, 8 and 12, on a $6 \times 6$
cluster.} 
\label{fig8}
\end{figure}

\noindent a function of hole doping.  As remarked before, many
experimental measurements of the specific heat for high temperature
cuprates are available. In general it is very difficult to separate the
electronic contribution to the specific heat in the normal state from
the phononic part.  Also many experiments have been performed in the
superconducting phase, where the existence of an intrinsic linear
contribution to the specific heat would indicate the absence of a gap
and thus non-conventional behavior.\cite{stupp,mori,collo} Since the
superconducting phase cannot be reached in QMC simulations, our results
will be compared with experiments performed in the normal state.  For
$La_{2-x}Sr_xCuO_4$ it was observed in Ref.\cite{loram} that the linear
term $\gamma$ of the specific heat in the normal phase increases with
doping between $x=0.12$ and 0.25. However, studies of the same material
performed later\cite{wada} showed that $\gamma$ increases with $x$ for
$x>0.1$ reaching a maximum value at optimal doping $x\approx 0.15$ and
then decreasing in the overdoped regime.  This behavior is in agreement
with the Van Hove scenario\cite{AFVH} where the density of states
reaches a maximum at optimal doping.  The behavior of $\gamma$ for a
metal-insulator transition was also studied for $Sr_{1-x}La_xTiO_3$ in
Ref.\cite{tokura}. They observed that $\gamma$ increases as the
transition is approached from the metallic side. Through the relation
$\gamma=m^* \gamma_0/m$, where $\gamma_0$ and $m$ are the linear
coefficient and the mass for free electrons, it was found that the
effective mass of the quasiparticles $m^*$ increases as the transition
is approached. Loram et al.\cite{loram2} studied $\gamma$ as a function
of doping at $T=280K$ in $YBa_2Cu_3O_{6+x}$. $\gamma$ appears to
increase with doping reaching a plateau for $x \approx 0.45$.

The first step to study numerically the specific heat at finite density
is to analyze the finite size effects. They are stronger than at
half-filling, but still moderate as can be seen in Fig.7 where $E$ vs
$T$ for $U=0$, 4, 8 and 12 at $\langle n \rangle =0.75$ on $4 \times 4$
and $6 \times 6$ clusters is shown. Away from half-filling it was very
difficult to obtain accurate results on $8 \times 8$ clusters at low
temperature due to the well-known sign problem.  However, since FSE are
stronger in weak coupling and we have observed that for $U=0$, where
results can be obtained exactly, there is only a

\begin{figure}[h]
\centerline{\psfig{figure=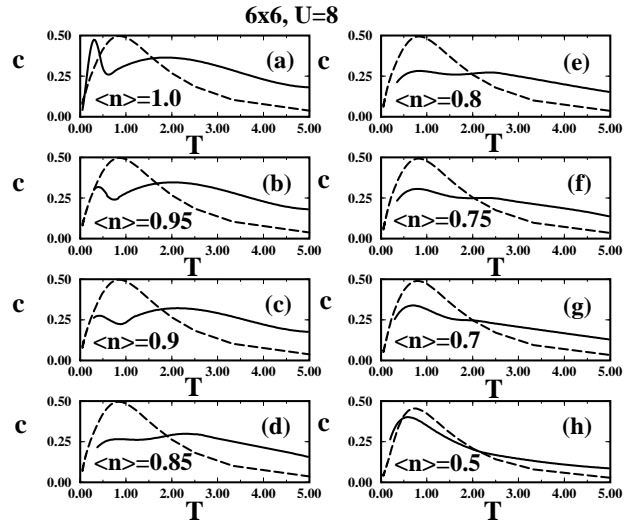,width=8cm,bbllx=55pt,bblly=36pt,bburx=610pt,bbury=700pt,angle=270}}
\caption{$c$ vs $T$ for $U=8$ (solid line) on a $6 \times 6$ cluster at
different densities. The dashed line indicates results for $U=0$
obtained using a $200 \times 200$ cluster.}
\label{fig9}
\end{figure}

\noindent small difference between the $6 \times 6$ and $8 \times 8$
results, then as before, $6 \times 6$ lattices were used in our studies
away from half-filling.

In Fig.8.a the specific heat as a function of $T$ at ${\langle n \rangle
}=0.75$ for different values of $U$ is presented.  It can be observed
that the spin peak is substantially reduced compared with the results at
half-filling, but it is still present in strong coupling for $U=8$ and
12 indicating the existence of short range antiferromagnetic
correlations.  In weak coupling, i.e. for $U=4$, the spin feature has
disappeared, and the curve is similar to the non-interacting one.  The
specific heat increases in the region where the minimum between the two
peaks existed at half-filling.  At quarter filling (Fig.8.b), $c$ has a
behavior that resembles free electrons independently of the value of
$U$. Thus, here the electrons are approximately weakly interacting at
all couplings.

Let us consider in more detail the special case of $U$=8.  This value of
the coupling was selected since according to calculations of the optical
conductivity it is suitable to reproduce some normal state experimental
results\cite{rev}.  In Fig.9 the specific heat as a function of
temperature is presented for different values of the density $\langle n
\rangle$. The continuous line indicates the results for $U=8$ on a $6
\times 6$ cluster while the dashed line denotes the non-interacting
$U=0$ results on a $200 \times 200$ lattice. Such a large cluster in the
non-interacting case was used to avoid finite size effects which are
strong in this limit at the low temperatures where the linear behavior
occurs.  Again it should be remarked that this problem occurs in weak
coupling at very low $T$ and, thus, our $U=8$ results are not expected
to be contaminated by size effects.  In Fig.9 it can be seen that for
$U=8$ the intensity of the spin peak decreases smoothly with doping.  At
10\% hole doping (i.e.  $\langle n \rangle = 0.90$) its intensity
diminishes by 40\%, a result in agreement with Ref.\cite{heis} where the
$t-J$ model was studied.  Note that for $\langle n \rangle \sim 0.8$ the
specific heat is almost

\begin{figure}[h]
\centerline{\psfig{figure=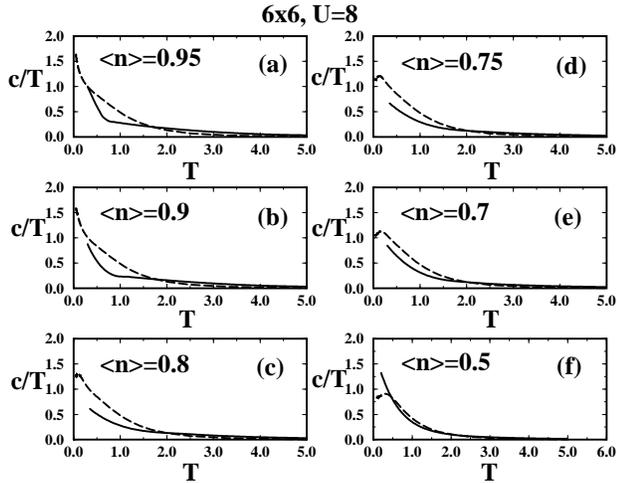,width=8cm,bbllx=50pt,bblly=36pt,bburx=610pt,bbury=750pt,angle=270}}
\caption{$c/T$ vs $T$ for $U=8$ (solid line) on a $6 \times 6$ cluster
at different densities. The dashed lines denote results for $U=0$
obtained using a $200 \times 200$ cluster.} 
\label{fig10} 
\end{figure}

\noindent flat in a broad range of temperatures. Here it
is difficult to resolve the spin and charge peaks from the data. We
expect that at this density or lower the spin correlations are no longer
important, even those of short-range, in agreement with previous
spectral function studies performed in the Hubbard model.\cite{dyn}
Reducing further the density from $\langle n \rangle =$0.75 to 0.5 a
single peak structure that resembles the non-interacting specific heat
curve becomes dominant.

An important issue in this context is the calculation of couplings and
densities where the system changes from insulator to metal.  Metallic
behavior is characterized in the specific heat by the existence of a
linear coefficient $\gamma$. In two dimensions it was found
that\cite{bedell}

$$
c\approx \gamma T+\Gamma_{2D}T^2+...,
\eqno(2)
$$

\noindent with $\Gamma_{2D}$ positive in strong coupling. 

The experimentalists often present plots of $c/T$ vs $T^2$ when
addressing $\gamma$. Analogously, in Fig.10 the continuous line denotes
$c/T$ vs $T$ for $U=8$ at different densities, while the dashed line
indicates the non-interacting case. The lowest temperature that was
confidently reached in this study away from half-filling is $T=0.3$. It
is clear that this temperature is too high to observe the linear
behavior in $c/T$ since, according to Eq.(2), the slope of the curve has
to be positive at very low temperature.  Clearly, if the system behaves
as a Fermi liquid a maximum has to appear in the curve at a lower
temperature than reached in this study. The non-interacting results show
indeed the linear behavior at very low temperatures. However, note that
the value of $\gamma$ for non-interacting electrons is not much
different from the value of $c/T$ at the maximum in Fig.10 at all
densities.  Thus by extrapolating the $U=8$ curves to zero we expect to
obtain a good

\begin{figure}[h]
\centerline{\psfig{figure=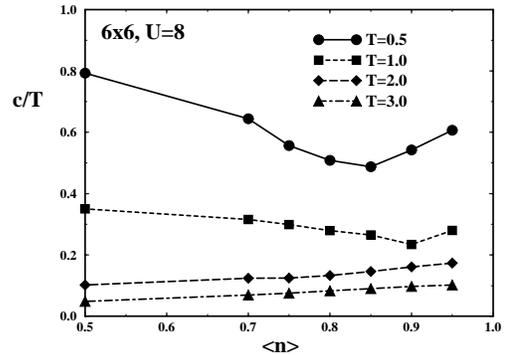,width=7cm,bbllx=98pt,bblly=36pt,bburx=590pt,bbury=750pt,angle=270}}
\caption{$c/T$ vs $\langle n \rangle$ for $U=8$ at different temperatures.}
\label{fig11}
\end{figure}

\noindent approximation to the value of $\gamma$.  However, since we
cannot reach lower temperatures the existence of anomalous non-Fermi
liquid behavior can certainly not be ruled out as remarked in
Ref.\cite{heis}.

In Fig.11, $c/T$ as a function of doping is presented at different
temperatures. Notice that the lowest temperature $T=0.5$ shown in Fig.11
 corresponds to $\sim 2000K$, if $t=0.4eV$ is used.  This is much higher
than $T=280K$ which is the highest temperature used in experiments.
However, for $T=0.5$ it was here observed that $c/T$ increases with
doping for $\langle n \rangle \leq 0.8$ in agreement with some
experimental results\cite{loram2}, and the same behavior is observed at
$T=1$. For higher $T$ the ratio $c/T$ increases for increasing density
$\langle n \rangle$.

\section{Summary}

The specific heat of the two dimensional Hubbard model has been
calculated for different couplings and electronic densities as a
function of temperature. At half-filling and as the coupling $U$
increases a low temperature peak associated with the spin degrees of
freedom moves to lower temperatures, while a high temperature feature
associated with the charge degrees of freedom moves to higher
temperatures. At very low temperatures $c \approx \delta T^2$ as
predicted by spin-wave theory and $\delta$ tends to the Heisenberg value
($\delta \approx 1.1$) for large coupling $U$. Away from half-filling we
observed that the spin feature weakens with doping, and it disappears
for $\langle n \rangle \leq 0.75$ working at $U=8$. This suggests the
absence of important antiferromagnetic correlations below that density.
We were not able to reach temperatures low enough to decide whether the
system is metallic or has anomalous behavior away from half-filling.
However, by evaluating $c/T$ we were able to make comparisons with
experimental results. At the lowest temperatures that we could reach we
found that $c/T$ increases with hole doping for $\langle n \rangle <
0.9$. This behavior is similar to experimental results for
$YBa_2Cu_3O_{6+x}$.\cite{loram2}

\section{Acknowledgements}

We thank E.~Miranda, K.~Bedell, S.~von Molnar and E.~Dagotto for useful
conversations.  A.M. is supported by NSF under grant DMR-95-20776.
Additional support is provided by the Office of Naval Research under
grant N00014-93-0495, the National High Magnetic Field Lab and MARTECH.
We thank ONR for providing access to their Cray-YMP and CM5
supercomputers.
 
\vfil\eject

\end{document}